# Reversible long-range domain wall motion in an improper ferroelectric

*M. Zahn[1,2], A.M. Müller[3], K. P. Kelley[4], S. M. Neumayer[4], S. V. Kalinin[5], I. Kézsmarki[2], M. Fiebig[3], Th. Lottermoser[3], N. Domingo[4], D. Meier[1,\*], J. Schultheiß[1,6,\*]*

[1] Department of Materials Science and Engineering, Norwegian University of Science and Technology (NTNU), 7034, Trondheim, Norway
[2] Center for Electronic Correlation and Magnetism, Institute of Physics, University of Augsburg, Germany
[3] Department of Materials, ETH Zurich, 8093 Zurich, Switzerland
[4] Center for Nanophase Materials Science, Oak Ridge National Laboratory, Oak Ridge, USA
[5] Department of Materials Science and Engineering, University of Tennessee, Knoxville, USA
[6] Department of Mechanical Engineering, University of Canterbury, 8140 Christchurch, New Zealand
\* dennis.meier@ntnu.no, jan.schultheiss@ntnu.no

**Reversible ferroelectric domain wall movements beyond the 10 nm range associated with Rayleigh behavior are usually restricted to specific defect-engineered systems. Here, we demonstrate that such long-range movements naturally occur in the improper ferroelectric ErMnO$_3$ during electric-field-cycling. We study the electric-field-driven motion of domain walls, showing that they readily return to their initial position after having travelled distances exceeding 250 nm. By applying switching spectroscopy band-excitation piezoresponse force microscopy, we track the domain wall movement with nanometric spatial precision and analyze the local switching behavior. Phase field simulations show that the reversible long-range motion is intrinsic to the hexagonal manganites, linking it to their improper ferroelectricity and topologically protected structural vortex lines, which serve as anchor point for the ferroelectric domain walls. Our results give new insight into the local dynamics of domain walls in improper ferroelectrics and demonstrate the possibility to reversibly displace domain walls over much larger distances than commonly expected for ferroelectric systems in their pristine state, ensuring predictable device behavior for applications such as tunable capacitors or sensors.**

Ferroelectrics have a switchable spontaneous polarization which is used, for example, in memory applications,[1] and their large piezoelectric and dielectric responses are essential for capacitors, sensors and actuators.[2, 3] Proper ferroelectrics, whose primary order parameter is the spontaneous polarization, such as Pb(Zr,Ti)O$_3$ and BaTiO$_3$, exhibit a qualitatively well-defined electric-field-driven polarization reversal process[4] that is determined by the elastic and electrostatic boundary conditions[5, 6]. At the local scale, the reversal is governed by domains, following material-dependent nucleation and growth processes that involve the movement of ferroelectric domain walls.[7] In general, for small electric fields reversible domain wall movements can occur (Rayleigh's law), whereas larger electric fields lead to nonlinearity, manifesting as Barkhausen jumps.[8] Characteristic distances reported for reversible ferroelectric domain wall movements according to the Rayleigh's law are usually in the sub-10-nm range.[9] To increase the regime in which reversible domain wall movements occur, different approaches have been tested, including point defect[10-12] and domain engineering,[13] as well as the cycling of ferroelectric domain walls between pinning centers,[14] which provide a strong restoring force to the domain walls. Following these approaches, electric-field-driven reversible domain wall motions over distances up several micrometers have been achieved. In contrast to proper ferroelectrics, the local dynamics and reversible displacements of domain walls in improper ferroelectrics, where the electric polarization arises as a secondary order parameter,[15, 16] remain relatively unexplored.

Here, we investigate the domain wall dynamics in the improper ferroelectric model system ErMnO$_3$. By applying band-excitation piezoresponse force microscopy (BE-PFM),[17] we map the ferroelectric domains and resolve their local response to applied electric fields. We find that away from the topologically protected structural vortex lines, which are characteristic for the domain pattern in hexagonal manganites,[18-20] domain walls readily move under the electric field. Most intriguingly, the domain walls consistently return to their original positions after bipolar electric-field cycling, even after being displaced by distances exceeding 250 nm. Complementary phase field simulations exhibit the same switching behavior, indicating that this phenomenon is characteristic for hexagonal manganites and linked to their improper ferroelectric order.

ErMnO$_3$ is a geometrically driven improper ferroelectric, where the primary order parameter relates to a structural trimerization mode. It's fundamental domain[19, 20] and domain-wall physics[21, 22] are well-understood, making it an ideal model system for dynamical investigations at the nanoscale. The system is a uniaxial ferroelectric with a spontaneous polarization, $P \approx 6$ µC/cm², parallel to the hexagonal $c$-axis (space group $P6_3cm$). The domain structure consists of ferroelectric 180° domains that meet at topologically protected six-fold structural vortex lines, which from at the improper ferroelectric phase transition as described elsewhere.[23] The family of hexagonal manganites is attracting broad attention due to its unusual physical properties, including multiferroicity,[24, 25] functional domain walls,[22, 26] Kibble-Zurek scaling,[27] and inverted domain-size / grain size scaling,[28, 29] with potential applications in barrier layer capacitors[30]. In this work, we use polycrystalline ErMnO$_3$ from the same batch as studied in ref. [28], where the interested reader can find relevant details concerning the processing conditions of the sample and a crystallographic and microstructural characterization.

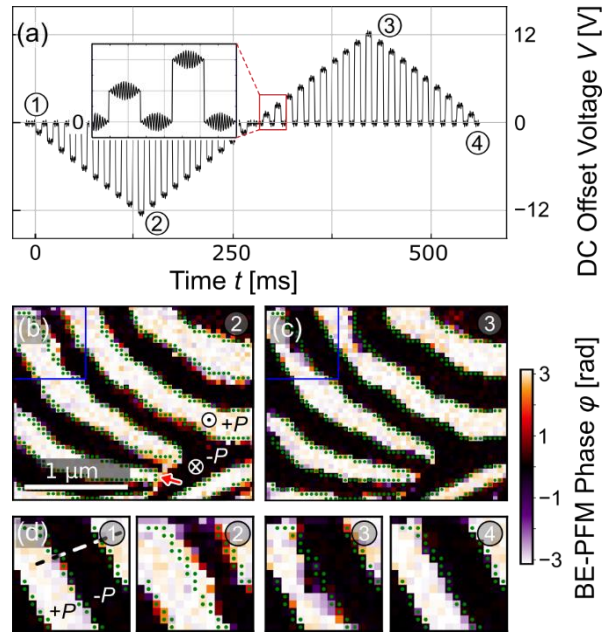

**Figure 1: BE-PFM data showing the voltage-dependent domain-structure evolution of ErMnO$_3$.** a) Time-dependence of the applied DC bipolar voltage sequence, corresponding to a square triangular function. The inset schematically visualizes the BE-PFM measurement applied to monitor the piezoresponse around the tip-sample resonance frequency. Corrected off-field BE-PFM phase images of the domain structure at b) maximum negative and c) maximum positive voltage. White areas represent domains with upward polarization, +P, whereas black areas indicate downward polarization, −P. The red arrow indicates the position of the intersection of a vortex line with the surface. Green dots indicate the position of the ferroelectric domain wall prior to the application of the bipolar volage sequence (at $t$ = 0 ms). A corresponding BE-PFM amplitude image is displayed in Figure S1. The continuous change of the ferroelectric domain structure is displayed in a movie in the Supplementary Information (Supplementary Movie 1). A zoom-in into the region marked by the blue square, featuring its voltage-dependent domain-structure evolution, is displayed in d). The time-dependent BE-PFM phase (after the application of the DC pulses) is extracted along the dashed white line and displayed in Figure 3d. The dimension of the square is 0.66 x 0.66 µm².

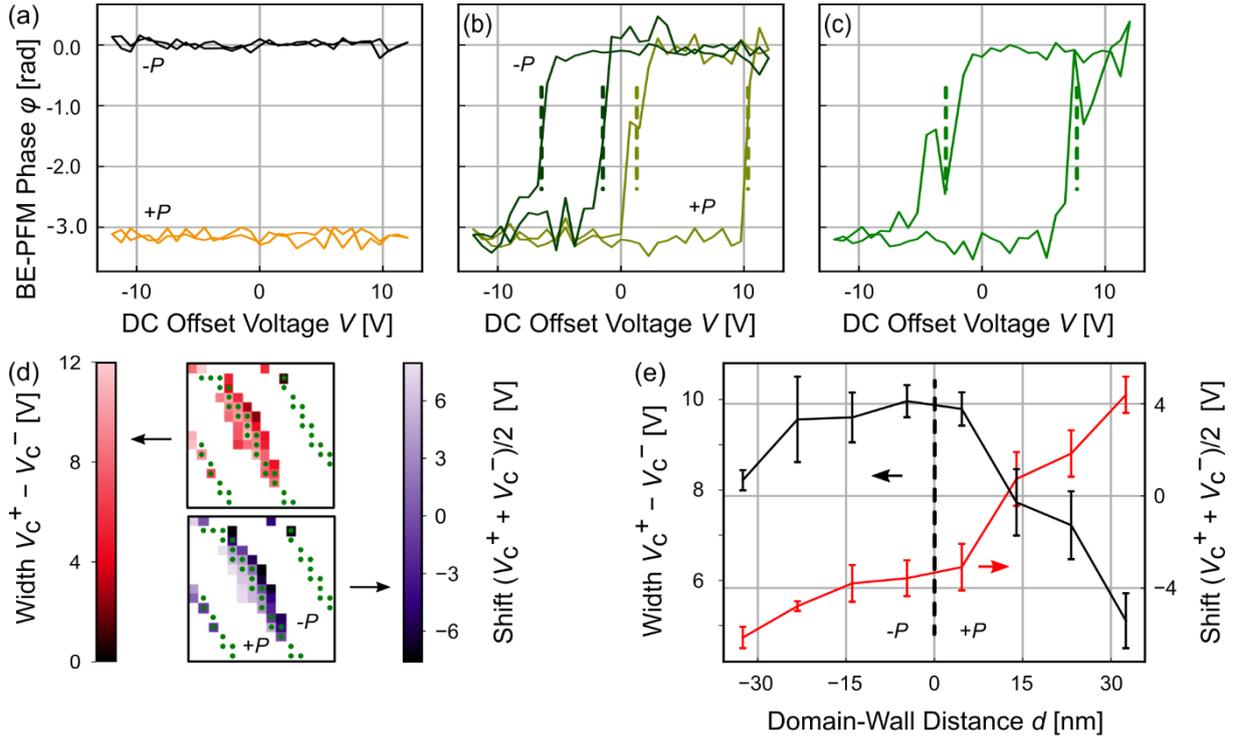

**Figure 2: Nanometric evaluation of the switching behavior in ErMnO$_3$.** The voltage-dependent BE-PFM phase response illustrates distinct switching types relative to the distance from the ferroelectric domain wall: a) the BE-PFM phase response is constant for all DC offset voltages and no switching is observed at distances larger than 150 nm from the domain wall. The -$P$ domains exhibit a voltage-independent phase of 0, whereas the +$P$ domains exhibit a phase of $-\pi$. b) Asymmetric hysteresis loop measured at two different sites of the domain wall at a distance of ~30 nm. The light green loop represents the response measured in a +$P$ domain and the dark green loop corresponds to the response measured in a -$P$ domain. c) Symmetric hysteresis observed at the position of the ferroelectric domain wall. Dashed lines in b) and c) represent the positive and negative coercive voltage $V_c^+$ and $V_c^-$, respectively. Panel d) illustrates the width and horizontal shift of the hysteresis loop, showcasing the validity of our classification (panel a to c) for the area of interest (Figure 1d). Green dots indicate the position of the ferroelectric domain wall prior to the application of the bipolar voltage sequence ($t$ = 0 ms). e) Hysteresis shift and width, assessed in the vicinity of the domain wall in the section shown in Figure 2d, depicted as a function of the distance from the domain wall at time $t$ = 0 ms (displayed as a black dashed line).

We begin by measuring the local electric-field response using BE-PFM[17, 31] as presented in Figure 1. In contrast to conventional PFM, which operates at a single frequency, the response is detected in BE-PFM within a specified frequency range (here 360 to 420 kHz), allowing to capture the broad mechanical resonance frequency of the tip-sample system.[32] Figure 1a shows a schematic illustration of the measurement principle. To image the polarization-reversal process, a bipolar triangular voltage signal sequence (Figure 1a), regularly used in SS-PFM,[33] is applied point by point to the surface via the probe tip. The BE-PFM response is measured after pulse application, as shown in the inset to Figure 1a (the protocol applied to correct for electrostatic background contributions and electric-field screening effects[34, 35] is described in Supplementary Figure S1 and S2). The initial vertical BE-PFM phase response of our polycrystalline ErMnO$_3$ specimen is displayed in Figure 1b (for corresponding amplitude data see Supplementary Figure S3). We find that the BE-PFM response is maximized in the vertical channel, which leads us to the conclusion that the grain has primarily an out-of-plane polarization component, i.e., the $c$-axis is approximately perpendicular to the imaged surface. Ferroelectric ±$P$ domains can be distinguished based on their characteristic BE-PFM phase response, with -$P$ domains showing a phase shift of 0 rad and +$P$ domains showing a -$\pi$ rad phase shift, consistent with their opposite polarization directions. The BE-PFM data shows the well-established domain structure of ErMnO$_3$,[18, 19, 22] consisting of ferroelectric 180° domains which come together by forming a characteristic sixfold line defect. The surface intersection point of the line is marked by the red arrow in Figure 1b.

Representative BE-PFM phase images recorded at maximum negative (−12.0 V) and positive (+12.0 V) voltages are displayed in Figure 1b and c, respectively. We find pronounced spatial variations in the response of the ferroelectric domains depending on the local structure. For example, near the structural vortex lines, we do not resolve domain wall movements, corroborating that the vortices act as anchor points for the ferroelectric domain walls.[36, 37] In contrast, substantial changes in the domain wall position are observed away from the vortex lines as presented in Figure 1d. Figure 1d shows the voltage-dependence of the BE-PFM phase response for the area marked by the blue square in Figure 1b and c. To demonstrate the effect of the electric field on the ferroelectric domains, the initial domain wall position is highlighted by green dots. Under application of a negative voltage the −$P$ domains contract, followed by an expansion under positive voltage. Most interestingly for this work, we find that the initial and final domain-wall positions in Figure 1d are almost indistinguishable, indicating that the domain walls revert to their starting position after applying the bipolar electric-field cycle. The behavior is consistent over the sample as confirmed by Figure S4, which displays the analysis at a different position. Our results align with previous scanning electron microscopy studies, which demonstrate that the initial and final state after ion- or electron-beam irradiation are similar.[38] In contrast to those studies, we apply voltage via an electrical contact, enabling precise monitoring of domain wall positions under controlled conditions.

To understand the local polarization reversal, we examine the voltage-dependent BE-PFM phase response on the nanoscale in the region of interest (Figure 1d). The BE-PFM phase response can be classified as a function of the distance from the domain wall. Representative responses are displayed in Figure 2a-c. At distances exceeding 150 nm from the original domain wall position, the BE-PFM phase response is independent of the applied voltage pulse (Figure 2a), showing that the local polarization does not switch up to the maximum applied voltage of ±12.0 V. As we approach the domain wall, highly asymmetric hysteretic BE-PFM phase responses emerge as a function of the voltage, with the effect being position-dependent, varying between 40 nm and 150 nm from the initial domain wall position. The sign of the asymmetry in the hysteresis depends on the side from which the domain wall is approached (Figure 2b). For both sides, a change of the BE-PFM phase by $\pi$ rad under the maximum applied voltage can be measured, which is an indication of local polarization reversal. Depending on the side of the domain wall, the asymmetry in the hysteresis loop is expressed by the absence of a negative or positive coercive voltage. Mechanistically, the BE-PFM data indicates that polarization reversal occurs by the attraction or repulsion of the ferroelectric domain wall towards or away from the AFM tip,[39, 40] implying that the domain wall reverts back to its initial position after application of the bipolar electric-field cycle. Finally, at a position close to the ferroelectric domain wall (Figure 2c), a hysteresis loop develops as expected for electric-field-driven polarization reversal. The behavior can be quantified by measuring the width of the hysteresis loop, $V_c^+ - V_c^-$, and its shift, $0.5(V_c^+ - V_c^-)$. Voltages $V_c^+$ and $V_c^-$ are displayed as dashed lines in Figure 2b and c. As depicted in Figure 2d, the width of the hysteresis loop reaches up to ~12 V at the domain wall and gradually decreases away from it. Similarly, the hysteresis shift peaks ±6 V away from the domain wall and gradually decreases towards the position of the wall. As expected, the hysteresis shift is positive when approaching the domain wall from the +P domain and has a negative value when approaching the domain from the -P domain. Figure 2e displays the average hysteresis width alongside the hysteresis shift, plotted as a function of the distance from the domain wall position at $t = 0$ ms, as defined in Figure 1a (the position is indicated by a dashed straight line). The behavior is interesting as it shows that at the nanoscale, distinct switching behaviors can be achieved within the same electric field range, depending on the relative positioning of the electrode with respect to the domain walls and vortex lines.

To evaluate whether the response to the electric field is specific to the ErMnO$_3$ polycrystal or representative for the family of hexagonal manganites in general, we compare the experimental results with phase-field simulations.[41] Following the established phase-field model for hexagonal manganites, the system is described by a Landau expansion of the trimerization tilt amplitude $Q$, the azimuthal tilt phase $\Phi$, and the polar mode $\mathcal{P}$,[20, 42, 43] as elaborated in the method section. The model reproduces the characteristic domain structure of ErMnO$_3$, including the topologically protected vortex lines, as displayed in Figure 3a. Importantly, the data allows for investigating the impact of an electric field on different surfaces and within the bulk material, giving a 3D model of the electric-field-induced domain structure. For this purpose, we consider the coupling between the energy density and the electric field as an additional term in the free energy, $F = F_{\text{Landau}} - PE$, where $P$ is the polarization and $E$ is the external electric field, as explained in more detail in the method section.

The simulated domain structure for a triangular electric field sequence analogous to the experiment (Figure 1a) is displayed in Figure 3a and b, visualizing the electric-field driven evolution of the ferroelectric domain structure from the initial to the maximum electric field, respectively. The simulation reproduces the experimentally observed contraction and expansion of the ferroelectric domains in response to the applied electric field. Close to the vortex lines, variations in the domain structure are much less pronounced than elsewhere, corroborating that these lines act as anchor points for the domain walls[36, 37]. Interestingly, in agreement with our experimental observations, the simulation shows that the ferroelectric domain walls return to their initial position after application of the bipolar electric-field cycle. The latter is displayed by the electric-field-dependent evolution of two representative walls in Figure 3c. Green dots are used to indicate the position of the ferroelectric domain walls, showing how they evolve as function of the electric field. For comparison, experimental data extracted from Figure 1d are presented in Figure 3d. Different from the simulation, jump-like changes in position, exemplarily highlighted by white arrows, arise in the experiment, which we attribute to imperfections in the crystallographic structure of the material. For example, it has been calculated that oxygen interstitials have a lower formation energy at the domain walls and, hence, promote pinning.[26, 44] Despite such jump-like changes, experiment and theory qualitatively exhibit the same behavior. The domain walls relax back to their initial positions after the bipolar electric-field cycle. This behavior suggests that the reversibility in the domain switching is linked to the improper ferroelectric order of ErMnO$_3$, with the relaxation of the primary structural order parameter acting as an additional resorting force, overcoming the pinning potential of defects. Remarkably, the domain walls revert to their original position after having traversed distances of up to 284±25 nm (see Figure 1d), moving between displacements under maximum positive and negative voltage. In general, even larger distances may be expected if higher voltages are applied with the upper limit being constrained by the proximity of neighboring domain walls. Most importantly, ErMnO$_3$ shows this reversibility in its pristine state, that is, without the need for engineering by, e.g., defect dipoles[45] or point defects[12, 14]. To gain additional insight into the microscopic interactions that govern the domain wall motion in ErMnO$_3$, extended atomic-level calculations are required, which could be performed as a next step. Most importantly for the present study, the agreement between the BE-PFM measurements and the phase-field simulations reflects that the observed behavior is common to improper ferroelectric hexagonal manganites and not specific to the ErMnO$_3$ polycrystal under investigation. Furthermore, this finding highlights that chemical defects play a secondary role in this system.

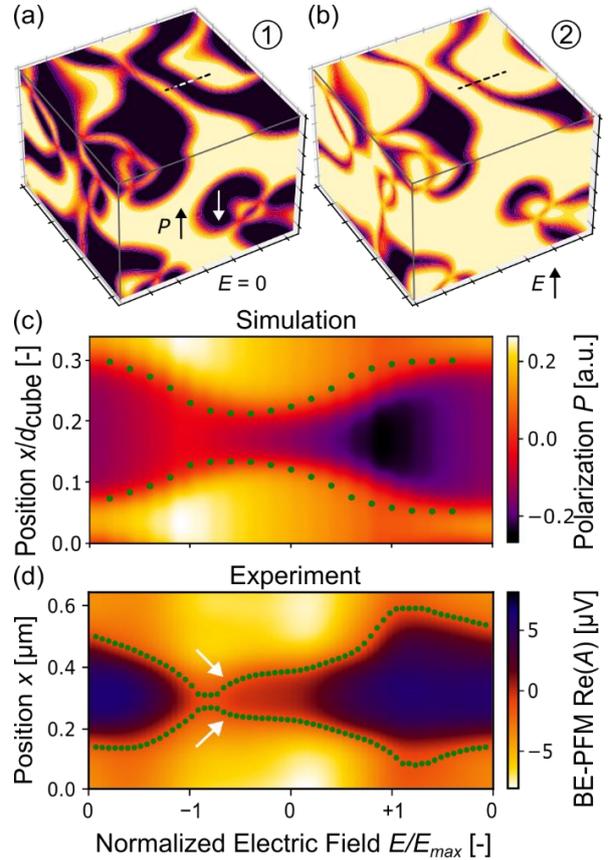

**Figure 3: Phase-field simulations showing the relation between the electric field and the domain structure in ErMnO$_3$.** The influence of an electric field on the ferroelectric domains is displayed for a) the initial configuration without an applied electric field and b) at the maximum negative electric field. The continuous change of the simulated ferroelectric domain structure is displayed in a movie in the Supplementary Material (Supplementary Movie 2). c) The polarization is extracted along the dotted line as a function of an equivalent triangular-shaped voltage sequence, analogous to our experiment (Figure 1a). d) Voltage-dependent data of the experiment BE-PFM phase extracted along a dotted line of the area of interest in Figure 1d is displayed. The green dots represent a guide to the eye for the electric-field-dependent positions of the domain wall, determined through thresholding. Jump-like changes in the position are highlighted by white arrows.

In conclusion, we have visualized the electric-field-driven local polarization reversal in improper ferroelectric ErMnO$_3$ using BE-PFM with nanometric precision. We attribute this phenomenon to the improper nature of the ferroelectric order, with the structural order parameter acting as a restoring force that promotes a relaxation of the domain walls to their initial position. Our observation of reversible long-range domain wall movements in ErMnO$_3$ is consistent with reports on the electric-field-induced contraction of domains into meandering lines[19, 21] and extends electric-field-dependent transmission electron microscopy studies[36] to the level of ferroelectric domains. We expect that the observed reversibility of the switching process is not limited to ErMnO$_3$ and the family of hexagonal manganites. Other possible candidate materials include isostructural ferrites,[46] indates,[47] and gallates[48], as well as hexagonal tungsten bronzes[49] or 2H compounds[50]. Furthermore, the absence of domain nucleation and growth offers precise control over the local switching process, rendering the proposed mechanism compelling for applications in tuneable capacitors, facilitating frequency agility and enhancing efficiency in communication systems.[51] A controllable domain wall position further ensures predictable device behavior, which is important for accuracy in sensing and actuating applications of ferroelectrics.[52]

**Methods**

**Synthesis and sample preparation.** Synthesis of ErMnO$_3$ powder was done by a solid-state reaction of dried Er$_2$O$_3$ (99.9% purity; Alfa Aesar, Haverhill, MA, USA) and Mn$_2$O$_3$ (99.0% purity; Sigma-Aldrich, St. Louis, MO, USA) powders. The powders were mixed and ball-milled (BML 5 witeg Labortechnik GmbH, Wertheim, Germany) for 12 hours at 205 rpm using yttria-stabilized zirconia milling balls of 5 mm and ethanol as dispersion medium. The reaction to ErMnO$_3$ was done by stepwise heating at 1000°C, 1050°C, and 1100°C for 12 hrs. More details on the powder processing can be found in ref. [28]. The powder was isostatically pressed into samples of cylindrical shape at a pressure of 200 MPa (Autoclave Engineers, Parker-Hannifin, Cleveland, OH, USA). Sintering was carried out in a closed alumina crucible at a temperature of 1350°C for 4 hours with a heating and cooling rate of 5 K/min (Entech Energiteknik AB; Ängelholm, Sweden).

**PFM measurements.** Prior to PFM measurements, the samples were lapped with a 9-µm-grained Al$_2$O$_3$ water suspension (Logitech Ltd., Glasgow, UK) and polished using silica slurry (SF1 Polishing Fluid, Logitech AS, Glasgow, Scotland). BE-PFM measurements were conducted with the same Cr/Pt coated Multi75E-G AFM probe (force constant, $k$≈3N/m, Budget Sensors, Sofia, Bulgaria), using a Cypher atomic force microscope (Oxford Instrument, Abingdon, UK). For BE-PFM imaging the electrical bias was applied to the surface via the probe tip, while the sample was grounded. More information on the measurement technique are provided in ref. [32]. The tip nominal radius was ~25 nm and the pixel size was ~47 nm. For the analysis of the data in Figs. 2 and 3, we took into account that the regular grid structure of the scan is non-collinear with the features of interest under investigation. BE-PFM was achieved by arbitrary-wave generator and data-acquisition electronics. A custom software program was employed to generate the probing signal and record local amplitude and phase hysteresis loops. BE-PFM measurements were performed at a frequency range of 360 to 420 kHz, a power spectral density of 0.57 V$^2$/kHz, and an amplitude of 5 V, respectively. Simple harmonic oscillator fits were subsequently applied to the measured spectra to extract amplitude and phase information at the resonance frequency. The SS-PFM triangular-square function had a maximum of ±12V, applied over 600 ms with a sequence of 64 steps of on-field and off-field measurements.

**Phase-field simulations.** Phase-field simulations were performed based on the Landau expansion of the free energy of ferroelectric hexagonal manganites as described in ref. [20] as

$$F_{Landau}(Q,\Phi,\mathcal{P}) = \frac{a}{2}Q^2 + \frac{b}{4}Q^4 + \frac{Q^6}{6}(c + c'\cos 6\Phi) - gQ^3\mathcal{P}\cos 3\Phi$$
$$-\frac{g'}{2}Q^2\mathcal{P}^2 + \frac{a_\mathcal{P}}{2}\mathcal{P}^2 + \frac{1}{2}\sum_{i=x,y,z}[s_Q^i(\partial_i Q \partial_i Q$$
$$+ Q^2\partial_i\phi\partial_i\Phi) + s_\mathcal{P}^i\partial_i\mathcal{P}\partial_i\mathcal{P}],$$

where $Q$ is the amplitude and $\phi$ is the angle of the lattice-trimerizing bipyramidal MnO$_5$ tilt, and $\mathcal{P}$ is a displacement field that corresponds to the polar mode $\Gamma_2^-$. This displacement field $\mathcal{P}$ is proportional to the polarization, $P$, of the system. To describe the coupling between the polarization, $P$, and the external electric field, $E$, an additional term $-PE$ is included. The parameters for the Landau expansion were chosen as in ref. [20], as $a = -2.626$ eVÅ$^{-2}$, $b = 3.375$ eVÅ$^{-4}$, $c = 0.117$ eVÅ$^{-6}$, $c' = 0.108$ eVÅ$^{-6}$, $a_P = 0.866$ eVÅ$^{-2}$, $g = 1.945$ eVÅ$^{-4}$, $g' = 9.931$ eVÅ$^{-4}$, $s_Q^z = 15.40$ eV, $s_Q^x = 5.14$ eV, and $s_\mathcal{P}^z = 52.70$ eV. To ensure stability, the gradient energy coefficient was set to $s_p^x = +8.88$ eV as validated in ref. [43]. Our system was simulated with a uniform Cartesian computational mesh with spacing $d_x = d_y = d_z = 0.2$ nm and $d_z = 0.3$ nm. Our simulations were performed on a mesh with size $n_x = n_y = n_z = 64$. Periodic boundary conditions were chosen in all three principal directions to simulate a bulk system. The Ginzburg-Landau equations were integrated with a Runge-Kutta 4 integrator with time steps of $\Delta t = 5 \cdot 10^{-4}$. The system was initialized with random values of all three order parameters. The initial domain pattern was then generated by evolving the system for $8 \cdot 10^4$ time steps, with no external electric field applied, i.e., $E = 0$. From this initial domain pattern, the system was integrated with an external field applied parallel to the polarization direction, i.e.,

$$E(n) = \begin{cases} \frac{4n}{N} \cdot E_{\max} & \frac{n}{N} \leq 1/4 \\ \left(1 - 4\left(\frac{n}{N} - \frac{1}{4}\right)\right) \cdot E_{\max} & \frac{1}{4} < \frac{n}{N} \leq \frac{3}{4} \\ \left(-1 + 4\left(\frac{n}{N} - \frac{3}{4}\right)\right) \cdot E_{\max} & \frac{3}{4} < \frac{n}{N} \end{cases}$$

which is similar to the experimentally applied bipolar voltage sequence displayed in Figure 1a, yet with a flipped sign of $E(n)$. Here, $N = 2 \cdot 10^4$ is the number of the time steps of a full cycle and $n$ is a given time step of the cycle. The maximal electric field has been chosen to $E_{\max} = 0.11$ V/Å.

**Acknowledgements**

J.S. acknowledges inspiring discussions with D. Damjanovic. O. W. Sandvik is acknowledged for initial analysis of the data. M.Z. acknowledges funding from the Studienstiftung des Deutschen Volkes via a doctoral grant and the State of Bavaria via a Marianne-Plehn scholarship J.S. and D.M. acknowledge NTNU Nano for the support through the NTNU Nano Impact fund and funding from the European Research Council (ERC) under the European Union's Horizon 2020 Research and Innovation Program (Grant Agreement No. 863691). J.S. acknowledges the support of the Alexander von Humboldt Foundation through a Feodor-Lynen research fellowship and the German Academic Exchange Service (DAAD) for a Post-Doctoral Fellowship (Short-term program). D.M. thanks NTNU for support through the Onsager Fellowship Program, the outstanding Academic Fellow Program. A.M.M. acknowledges funding from the Swiss National Science Foundation (SNSF) through grant numbers 200021_178825 and 200021_215423. The scanning probe microscopy research was supported by the Center for Nanophase Materials Sciences (CNMS), which is a US Department of Energy, Office of Science User Facility at Oak Ridge National Laboratory.

**Competing interests**

The authors declare no competing interests.

**Data availability statement**

The data that support the findings of this study are available from the corresponding author upon reasonable request.

**References**

[1] J. F. Scott, *Science* **2007**, *315*, 954.
[2] J. Schultheiß, G. Picht, J. Wang, Y. Genenko, L. Chen, J. Daniels, J. Koruza, *Prog. Mat. Sci.* **2023**, *136*, 101101.
[3] Q. M. Zhang, H. Wang, N. Kim, L. E. Cross, *J. Appl. Phys.* **1994**, *75*, 454.

[4] A. K. Tagantsev, L. E. Cross, J. Fousek, *Domains in ferroic crystals and thin films*, Springer, Heidelberg, Germany **2010**.
[5] Y. Ishibashi, Y. Takagi, *J. Phys. Soc. Jpn.* **1971**, *31*, 506.
[6] J. Schultheiß, L. Liu, H. Kungl, M. Weber, L. Kodumudi Venkataraman, S. Checchia, D. Damjanovic, J. E. Daniels, J. Koruza, *Acta Mater.* **2018**, *157*, 355.
[7] Y. A. Genenko, S. Zhukov, S. V. Yampolskii, J. Schütrumpf, R. Dittmer, W. Jo, H. Kungl, M. J. Hoffmann, H. von Seggern, *Adv. Funct. Mater.* **2012**, *22*, 2058.
[8] D. A. Hall, *J. Mater. Sci.* **2001**, *36*, 4575.
[9] H. Kronmuller, *Z. Angew. Phys.* **1970**, *30*, 9.
[10] L. Zhang, X. Ren, *Phys. Rev. B* **2005**, *71*, 174108.
[11] R. Ignatans, D. Damjanovic, V. Tileli, *Phys. Rev. Mater.* **2020**, *4*, 104403.
[12] V. Postnikov, V. Pavlov, S. Turkov, *J. Phys. Chem. Solids* **1970**, *31*, 1785.
[13] R. Ignatans, D. Damjanovic, V. Tileli, *Phys. Rev. Lett.* **2021**, *127*, 167601.
[14] T. Yang, V. Gopalan, P. Swart, U. Mohideen, *Phys. Rev. Lett.* **1999**, *82*, 4106.
[15] V. Dvořák, *Ferroelectrics* **1974**, *7*, 1.
[16] A. P. Levanyuk, D. G. Sannikov, *Sov. Phys. Uspekhi* **1974**, *17*, 199.
[17] S. Jesse, S. V. Kalinin, *J. Phys. D: Appl. Phys.* **2011**, *44*, 464006.
[18] M. Šafránková, J. Fousek, S. Kižaev, *Czech J. Phys.* **1967**, *17*, 559.
[19] T. Jungk, Á. Hoffmann, M. Fiebig, E. Soergel, *Appl. Phys. Lett.* **2010**, *97*, 012904.
[20] S. Artyukhin, K. T. Delaney, N. A. Spaldin, M. Mostovoy, *Nat. Mater.* **2014**, *13*, 42.
[21] T. Choi, Y. Horibe, H. T. Yi, Y. J. Choi, W. D. Wu, S. W. Cheong, *Nat. Mater.* **2010**, *9*, 253.
[22] D. Meier, J. Seidel, A. Cano, K. Delaney, Y. Kumagai, M. Mostovoy, N. A. Spaldin, R. Ramesh, M. Fiebig, *Nat. Mater.* **2012**, *11*, 284.
[23] M. Lilienblum, T. Lottermoser, S. Manz, S. M. Selbach, A. Cano, M. Fiebig, *Nat. Phys.* **2015**, *11*, 1070.
[24] T. Lottermoser, T. Lonkai, U. Amann, D. Hohlwein, J. Ihringer, M. Fiebig, *Nature* **2004**, *430*, 541.
[25] J. Schultheiß, L. Puntigam, M. Winkler, S. Krohns, D. Meier, H. Das, D. Evans, I. Kézsmárki, *Appl. Phys. Lett.* **2024**, *124*, 252902.
[26] J. Schaab, S. H. Skjærvø, S. Krohns, X. Y. Dai, M. E. Holtz, A. Cano, M. Lilienblum, Z. W. Yan, E. Bourret, D. A. Muller, M. Fiebig, S. M. Selbach, D. Meier, *Nat. Nanotechnol.* **2018**, *13*, 1028.
[27] S. M. Griffin, M. Lilienblum, K. T. Delaney, Y. Kumagai, M. Fiebig, N. A. Spaldin, *Phys. Rev. X* **2012**, *2*, 041022.
[28] J. Schultheiß, F. Xue, E. Roede, H. W. Ånes, F. H. Danmo, S. M. Selbach, L.-Q. Chen, D. Meier, *Adv. Mater.* **2022**, *34*, 2203449.
[29] K. Wolk, R. S. Dragland, E. C. Panduro, M. E. Hjelmnstad, L. Richarz, Z. Yan, E. Bourret, K. A. Hunnestad, C. Tzschaschel, J. Schultheiß, D. Meier, *Matter* **2024**, *7*, 3097.
[30] L. Zhou, L. Puntigam, P. Lunkenheimer, E. Bourret, Z. Yan, I. Kézsmárki, D. Meier, S. Krohns, J. Schultheiß, D. Evans, *Matter* **2024**.
[31] S. Jesse, R. Vasudevan, L. Collins, E. Strelcov, M. B. Okatan, A. Belianinov, A. P. Baddorf, R. Proksch, S. V. Kalinin, *Annu. Rev. Phys. Chem.* **2014**, *65*, 519.
[32] S. Jesse, S. V. Kalinin, R. Proksch, A. Baddorf, B. Rodriguez, *Nanotechnology* **2007**, *18*, 435503.
[33] S. Jesse, A. P. Baddorf, S. V. Kalinin, *Appl. Phys. Lett.* **2006**, *88*.
[34] T. Jungk, A. Hoffmann, E. Soergel, *J. Microsc.* **2007**, *227*, 72.
[35] N. Balke, S. Jesse, P. Yu, B. Carmichael, S. V. Kalinin, A. Tselev, *Nanotechnology* **2016**, *27*, 425707.
[36] M.-G. Han, Y. Zhu, L. Wu, T. Aoki, V. Volkov, X. Wang, S. C. Chae, Y. S. Oh, S.-W. Cheong, *Adv. Mater.* **2013**, *25*, 2415.
[37] K. Yang, Y. Zhang, S. Zheng, L. Lin, Z. Yan, J.-M. Liu, S.-W. Cheong, *Phys. Rev. B* **2017**, *96*, 144103.
[38] E. D. Roede, A. B. Mosberg, D. M. Evans, E. Bourret, Z. Yan, A. T. J. van Helvoort, D. Meier, *APL Mater.* **2021**, *9*, 021105.
[39] V. R. Aravind, A. N. Morozovska, S. Bhattacharyya, D. Lee, S. Jesse, I. Grinberg, Y. Li, S. Choudhury, P. Wu, K. Seal, *Phys. Rev. B* **2010**, *82*, 024111.
[40] A. N. Morozovska, S. V. Kalinin, E. A. Eliseev, V. Gopalan, S. V. Svechnikov, *Phys. Rev. B* **2008**, *78*, 125407.
[41] L.-Q. Chen, *Annu. Rev. Mater. Res.* **2002**, *32*, 113.
[42] A. Bortis, M. Trassin, M. Fiebig, T. Lottermoser, *Phys. Rev. Mater.* **2022**, *6*, 064403.
[43] F. Xue, N. Wang, X. Wang, Y. Ji, S.-W. Cheong, L.-Q. Chen, *Phys. Rev. B* **2018**, *97*, 020101.
[44] J. Schultheiß, J. Schaab, D. R. Småbråten, S. H. Skjærvø, E. Bourret, Z. Yan, S. M. Selbach, D. Meier, *Appl. Phys. Lett.* **2020**, *116*, 262903.
[45] L. X. Zhang, X. Ren, *Phys. Rev. B* **2005**, *71*.
[46] K. Du, B. Gao, Y. Wang, X. Xu, J. Kim, R. Hu, F.-T. Huang, S.-W. Cheong, *npj Quantum Mater.* **2018**, *3*, 1.
[47] S. Abrahams, *Acta Crystallogr. B* **2001**, *57*, 485.
[48] D. R. Småbråten, Q. N. Meier, S. H. Skjaervo, K. Inzani, D. Meier, S. M. Selbach, *Phys. Rev. Mater.* **2018**, *2*, 114405.
[49] J. A. McNulty, T. T. Tran, P. S. Halasyamani, S. J. McCartan, I. MacLaren, A. S. Gibbs, F. J. Lim, P. W. Turner, J. M. Gregg, P. Lightfoot, *Adv. Mater.* **2019**, *31*, 1903620.
[50] J. Varignon, P. Ghosez, *Phys. Rev. B* **2013**, *87*, 140403.
[51] Y. He, B. Bahr, M. Si, P. Ye, D. Weinstein, *Microsyst. Nanoeng.* **2020**, *6*, 8.
[52] P. Muralt, *J. Micromech. Microeng.* **2000**, *10*, 136.

**Supporting Material**

# Reversible long-range domain wall motion in an improper ferroelectric

M. Zahn[1,2], A.M. Müller[3], K. P. Kelley[4], S. M. Neumayer[4], S. V. Kalinin[5], I. Kézsmarki[2], M. Fiebig[3], Th. Lottermoser[3], N. Domingo[4], D. Meier[1,*], J. Schultheiß[1,6,*]

[1] Department of Materials Science and Engineering, Norwegian University of Science and Technology (NTNU), 7034, Trondheim, Norway

[2] Center for electronic correlation and magnetism, Institute of Physics, University of Augsburg, Germany

[3] Department of Materials, ETH Zurich, 8093 Zurich, Switzerland

[4] Center for Nanophase Materials Science, Oak Ridge National Laboratory, Oak Ridge, USA

[5] Department of Materials Science and Engineering, University of Tennessee, Knoxville, USA

[6] Department of Mechanical Engineering, University of Canterbury, 8140 Christchurch, New Zealand

1. Correction of BE-PFM signal for electro-mechanical contributions

In Figure S1a, the BE-PFM signal is displayed in a complex representation, visualizing the real and imaginary part[1] before the application of the bipolar triangular voltage signal sequence (indicated by 1) in Figure 1a. Reflecting the uniaxial ferroelectric nature of ErMnO$_3$, the data points are expected to cluster around two centers corresponding to the BE-PFM signal of the up- and down-polarized domains, respectively, forming a dumbbell-like closure. This general shape is observed for our data points. However, in addition a shift and a rotation of the closure with respect to the origin is observed, which can be attributed to electrostatic contributions. Figure S1b visualizes the BE-PFM signals in the complex plane for three positions: in +*P* and -*P* domains (orange and violet) and on a domain wall (green). As shown in Figure S1b, all three signals partially overlap. To correct for this effect, a parameterization is introduced as presented in Figure S1a. First, the two cluster centers for a given electric field are determined using *k*-means clustering with the number of the clusters, $k = 2$, on the complex BE-PFM data set, resulting in cluster centers visualized by red and green dots in Figure S1a. The middle in between them is referred to as the center of the dumbbell and is highlighted with a black dot. This data treatment now allows to describe the data set via two vectors, one from the origin to the center point of the dumbbell, $\vec{o}$, and one from one cluster center to the other, referred to as connector, $\vec{c}$.

Based on the two vectors, the BE-PFM signal is corrected so that the two cluster centers are projected on the positive and negative *Re*-axis, respectively, by first subtracting $\vec{o}$, followed by a subsequent multiplication by the normalized complex conjugate of the connector, $\vec{c}^*/|c|$, to compensate the rotation. The procedure to determine the cluster centers and correct both described contributions is repeated for every electric field separately. In Figure S1c, the signal evolution is displayed after correcting for the same three data points as in Figure S1b, which can now be clearly identified as belonging to oppositely oriented ferroelectric domains. It is evident that the orange and violet data points correspond to the oppositely oriented ferroelectric domains, whereas the green data set corresponds to a ferroelectric domain wall.

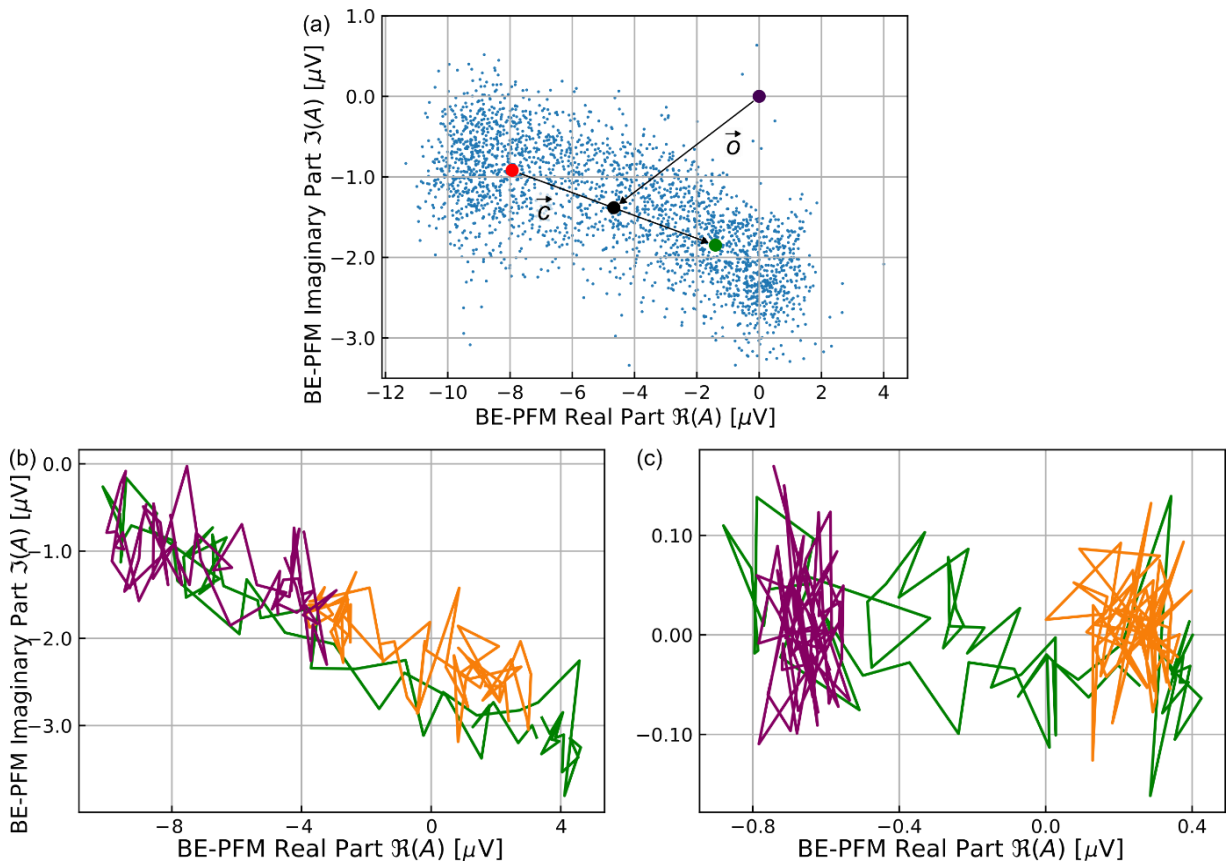

**Figure S1: Separation of ferroelectric and electro-mechanical signals.** a) The distribution of the BE-PFM signal in the complex plane before application of the bipolar triangular voltage signal sequence is of the expected dumbbell form. The dumbbell is displaced and rotated due to superimposed electromechanical contributions. To subtract for these electromechanical contributions, a parameterization with the vectors $\vec{o}$ (origin) and $\vec{c}$ (connector) is introduced. b) Evolution of the raw BE-PFM signal in the complex before compensating for the electro-mechanical contributions. Orange and violet data correspond to differently oriented domains, while green data points correspond to a domain wall. c) After compensating the electro-mechanical contributions, this becomes clear.

Figure S2a gives an overview of the evolution of the complex representation of the BE-PFM signal over the entire applied bipolar voltage sequence (Figure 1a). To analyze the evolution of the data systematically, we focus on the parameterization vectors introduced in Figure S1a. The trajectory of the center of the dumbbell captured by the origin vector, $\vec{o}$, is shown in Figure S2b. A clear voltage dependence of the origin vector is found, as outlined by the characteristic applied electric fields, 1-4. As explained in ref. [2], the difference in work function between tip and sample leads to a contact potential and an additional contribution to the captured signal, even in the absence of an external electric field. For our data set, this difference manifests in the displacement of the dumbbell from the origin of the complex plane. Since the BE-PFM measurement is performed in the absence of an electric field (see method section), a variation of the origin vector with electric field, as we observe in Figure S2b, is most likely a transient effect. This is supported by the fact that the found linear variation of the origin vector with the electric field is consistent with the theory introduced in ref. [2].

Next, the trajectory of the connection vector, $\vec{c}$, is displayed as amplitude and phase in Figure S2c. The phase is constant and independent of the applied electric field and can be calibrated to zero. The amplitude displays two local minima and maxima within the bipolar voltage profile sequence, indicating that the underlying effect couples quadratically to the applied electric field. According to the theory of electrostatic force microscopy,[3] free charge carriers, that screen the electric field, might accumulate or disperse for both direction of the field equally.

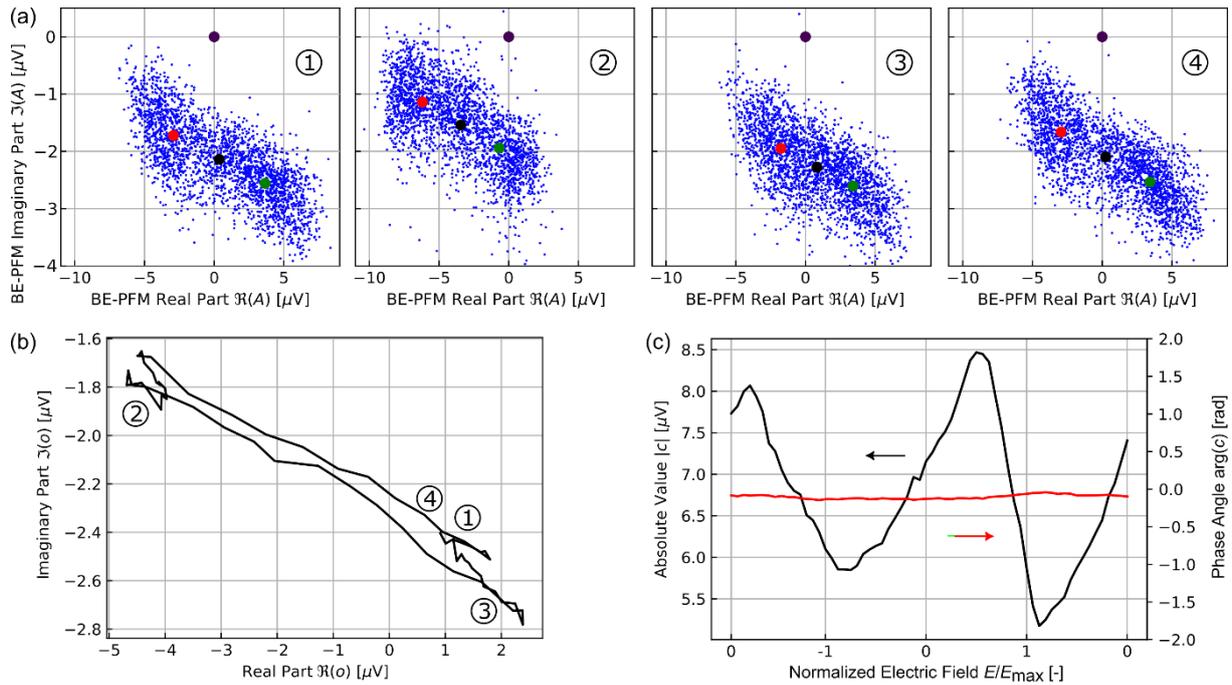

**Figure S2: Separation of ferroelectric switching and electro-mechanical signals over the bipolar voltage profile.** a) Evolution of the BE-PFM signal in the complex representation over the entire bipolar voltage signal sequence. The respective cluster centers described in Figure S2a are displayed. The numbers 1-4 are introduced in Figure 1a. b) Evolution of the extracted complex origin vector, $\vec{o}$, over the entire bipolar voltage signal sequence (Figure 1a). The applied voltage states are indicated by the numbers 1-4. c) Evolution of the connection vector, $\vec{c}$, as amplitude and phase as a function of the bipolar voltage signal sequence.

2. BE-PFM data showing ferroelectric order before voltage sequence application

Figure S3 shows the phase and amplitude to the BE-PFM signal. Both domains are separated by a $\pi$ rad BE-PFM phase difference with black domains pointing inwards and white domains outwards. The BE-PFM amplitude is equal in both domains, indicating a non-zero piezoelectric response within the domains, while a lower contribution is observed at the ferroelectric domain walls, indicating that the piezoelectric contribution at the ferroelectric domain walls vanishes[4].

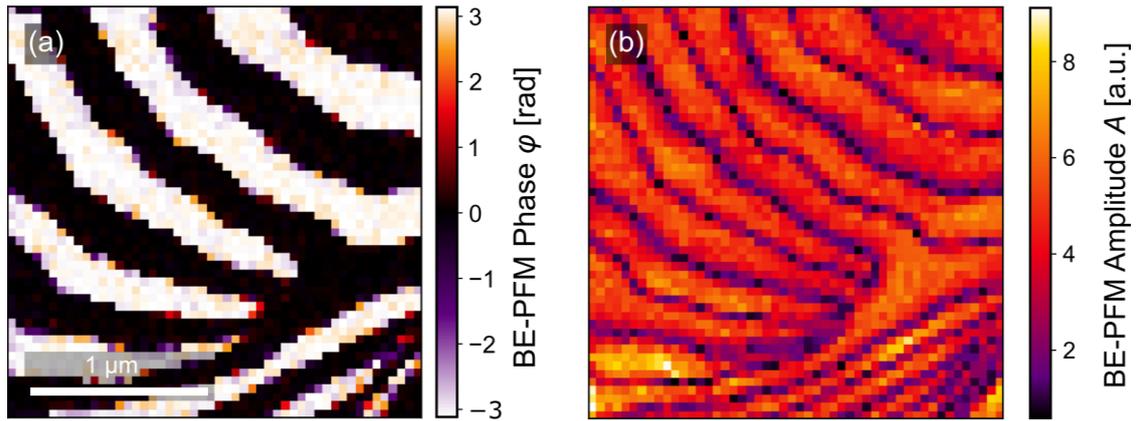

**Figure S3: BE-PFM data showing the domain structure before the application of bipolar triangular voltage signal sequence.** Corrected phase and amplitude data are displayed in a) and b), respectively. The phase data verifying the close to a π rad phase shift between the oppositely oriented domains. The amplitude data displays a clear minimum on every domain wall independent of its orientation, indicating a vanishing of the piezoresponse at the domain wall.

3. Evolution of the domain structure recorded at a second position

BE-PFM data has been recorded at different positions with the same bipolar electric field sequence. In addition to the data presented in the main text, a second data set recorded at a different position on an out-of-plane oriented grain is presented in Figure 1a. Similarly to Figure 1b-d, a coherent contraction and expansion of domains is observed. Again, the domain wall displacement is most pronounced away from the structural vortex lines, which can be for example observed in the stripe-like domains in the upper left corner of the BE-PFM image sequence.

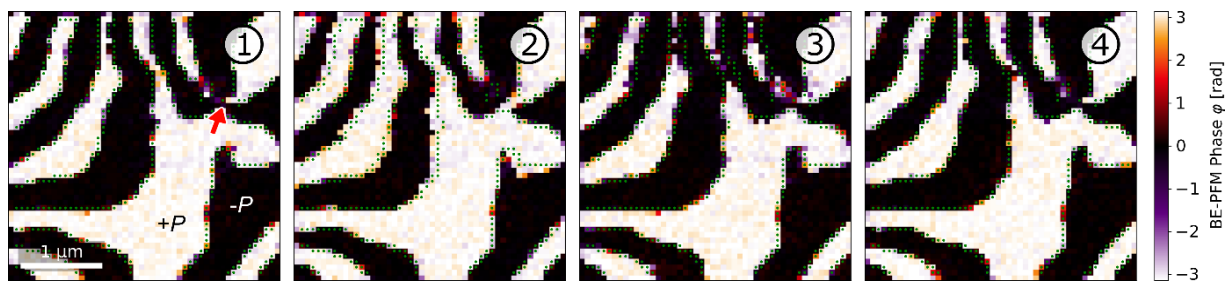

**Figure S4: Evolution of the BE-PFM phase recorded at a second position on the same material under a bipolar voltage sequence.** The numbers 1-4 correspond to applied electric fields as introduced in Figure 1a. White areas represent upwards-oriented domains, +P, and black areas represent downwards-oriented, -P, domains. The red arrow indicates the position of the intersection of a vortex line with the surface. Green dots highlight the initial positions of the ferroelectric domain walls, showing their evolution with applied voltage. Similar to Figure 1, near the structural vortex lines, we do not resolve domain wall movements. A notable reversibility of the domain structure is observed, for example in the stripe-like domains in the upper left corner. A movie

showing the electric-field dependent evolution of the BE-PFM phase and amplitude is provided in Supplementary Movie 3.

4. Movies, visualizing the electric-field driven evolution of the domain structure in experiment and simulation.

**Supplementary Movie 1:** BE-PFM amplitude and phase images showing the evolution of the ferroelectric domain structure during the electric field application, following the bipolar triangular voltage signal sequence of Figure 1a.

**Supplementary Movie 2:** Simulations showing polarization images obtained from phase field simulations. The data documents the evolution of the ferroelectric domain structure during the electric field application.

**Supplementary Movie 3:** BE-PFM amplitude and phase evolution analog to Supplementary Movie 1 for the position evaluated in Figure S4.

**References**

[1]    T. Jungk, A. Hoffmann, E. Soergel, *J. Microsc.* **2007**, *227*, 72.
[2]    N. Balke, S. Jesse, P. Yu, B. Carmichael, S. V. Kalinin, A. Tselev, *Nanotechnology* **2016**, *27*, 425707.
[3]    J. Schaab, A. Cano, M. Lilienblum, Z. Yan, E. Bourret, D. Meier, R. Ramesh, M. Fiebig, *Adv. Electron. Mater.* **2016**, *2*, 1500195.
[4]    E. Soergel, *J. Phys. D Appl. Phys.* **2011**, *44*, 464003.